\documentclass[superscriptaddress,
reprint,longbibliography,
preprintnumbers,nofootinbib,
amsmath,amssymb]{revtex4-1}

\pdfoutput=1
\usepackage{graphicx}
\usepackage{amsmath}
\usepackage{gensymb}
\usepackage{float}
\usepackage{physics}
\usepackage{xcolor}

\begin{document}

\raggedbottom

\title{Universal lower bounds on energy and momentum diffusion in liquids}

\author{K. Trachenko}
\affiliation{School of Physics and Astronomy, Queen Mary University of London, Mile End Road, London, E1 4NS, UK}
\author{M. Baggioli}
\affiliation{Instituto de Fisica Teorica UAM/CSIC, c/ Nicolas Cabrera 13-15, Cantoblanco, 28049 Madrid, Spain}
\author{K. Behnia}
\affiliation{Laboratoire de Physique et Etude des Mat\'eriaux (CNRS-Sorbonne Universit\'e-ESPCI), PSL Research University, 75005 Paris, France}
\author{V. V. Brazhkin}
\affiliation{Institute for High Pressure Physics, RAS, 108840, Troitsk, Moscow, Russia}

\begin{abstract}
Thermal energy can be conducted by different mechanisms including by single particles or collective excitations. Thermal conductivity is system-specific and shows a richness of behaviors currently explored in different systems including insulators, strange metals and cuprate superconductors. Here, we show that despite the seeming complexity of thermal transport, the thermal diffusivity $\alpha$ of liquids and supercritical fluids has a lower bound which is fixed by fundamental physical constants for each system as $\alpha_m=\frac{1}{4\pi}\frac{\hbar}{\sqrt{m_em}}$, where $m_e$ and $m$ are electron and molecule masses. The newly introduced elementary thermal diffusivity has an absolute lower bound dependent on $\hbar$ and the proton-to-electron mass ratio only. We back up this result by a wide range of experimental data. We also show that theoretical minima of $\alpha$ coincide with the fundamental lower limit of kinematic viscosity $\nu_m$. Consistent with experiments, this points to a universal lower bound for two distinct properties, energy and momentum diffusion, and a surprising correlation between the two transport mechanisms at their minima. We observe that $\alpha_m$ gives the minimum on the phase diagram except in the vicinity of the critical point, whereas $\nu_m$ gives the minimum on the entire phase diagram.
\end{abstract}

\maketitle

\section{Introduction}

Thermal energy can propagate by radiation, convection and conduction. The latter phenomenon refers to the travel of heat in matter in the absence of particle flow. Thermal energy can be carried by phonons and electronic quasi-particles in solids and liquids or molecular collisions in gases \cite{ashcroft,gases}. Although the two mechanisms of heat transfer by collective excitations or particles are conceptually simple, they can interestingly interact with other processes and give rise to a rich variety of behaviors. This is currently explored in a variety of materials including insulators, strange metals and cuprate superconductors, where new mechanisms are invoked to explain the experimental data (see, e.g., Refs. \cite{zaanen2,hartnoll1,behnia1,behnia2}). More recently, bounds on thermal conductivity and other properties were discussed, with the view that identifying and understanding these bounds is important for fundamental physics, predictions for theory and experiment as well as searching and rationalizing universal behavior \cite{zaanen2,hartnoll1,behnia1,behnia2,kss,hartnoll2,spin1,zaanen1,Blake:2016wvh,Baggioli:2020ljz,Grozdanov:2020koi}. These bounds are based on uncertainty relations and limits due to quantum physics.

Thermal conductivity is defined by the static Fourier equation, $J_Q=\kappa\, \frac{\partial T}{\partial x}$, where J$_Q$ is the heat current density and $ \frac{\partial T}{\partial x}$ is the temperature gradient in the $x$ direction. This equation is the thermal counterpart of the Ohm equation and defines $\kappa$ as a linear response to a static temperature gradient. Thermal diffusivity is described by the heat equation \cite{Joseph,Lawler}:

\begin{equation}
\frac{\partial T}{\partial t}= \alpha\, \frac{\partial^2T}{\partial x^2}
\label{Heat}
\end{equation}

\noindent where $\alpha=\frac{\kappa}{\rho c_p}$ is thermal diffusivity, $\rho$ is density and $c_p$ is heat capacity per mass unit. $\alpha$ plays the role of the diffusion constant quantifying the propagation of thermal energy.

The transport coefficients $\kappa$ and $\alpha$ vary in a wide range and depends strongly on the system, temperature and pressure. Here, we consider $\alpha$ in liquid and supercritical states of matter and show that despite these variations, $\alpha$ at its {\it minimum}, $\alpha_m$, universally attains a value

\begin{equation}
\alpha_m=\frac{1}{4\pi}\frac{\hbar}{\sqrt{m_em}}
\label{eq2}
\end{equation}

\noindent where $m_e$ and $m$ are electron and molecule masses, and back up this result by experimental data.

We subsequently introduce the elementary thermal diffusivity $\iota=\alpha_mm$, similarly to the elementary viscosity \cite{sciadv}, with the universal minimum set by fundamental constants as

\begin{equation}
\iota_m=\frac{\hbar}{4\pi}\left({\frac{m_p}{m_e}}\right)^{\frac{1}{2}}
\label{eq3}
\end{equation}

\noindent where $m_p$ is the proton mass.

We finally show that the theoretical minima of thermal diffusivity coincide with the minima of a physically distinct quantity, the kinematic viscosity $\nu_m$ discussed recently \cite{sciadv} and that the experimental ratio $\nu_m/\alpha_m$ is close to 1 and is in the range 0.4-1.7. Fundamentally, this closeness can be explained by observing that both $\alpha$ and $\nu$ at their minima are governed by the ``ultraviolet'' (UV) properties such as Bohr radius and Debye frequency. This suggests a wider universality of properties at their fundamental limit. We finally observe that (a) $\nu_m$ gives the minimum on the entire phase diagram of matter and (b) $\alpha_m$ gives the minimum on the phase diagram except in the vicinity of the critical point.

It is notable that the universal results (\ref{eq2}) and (\ref{eq3}) fixing the minimum for each system apply to the liquid state. Indeed, liquid properties are considered to be system-specific because interactions are strong and depend on the system. This circumstance is viewed to disallow a possibility of calculating liquid properties in general form \cite{landau}. A fundamental problem of liquid description is related to the absence of a small parameter \cite{ropp}: interactions and atomic displacements in liquids are both large, and this combination precludes using theories developed for gases and solids. For example, the theoretical calculation and understanding of liquid energy and heat capacity has remained a long-standing problem in research and teaching \cite{granato}, and started to lift only recently when new understanding of collective excitations in liquids came in \cite{ropp}. For these reasons, there is no tractable microscopic theory of thermal conductivity in liquids \cite{bird}. In view of these problems, the existence of universal bound for $\alpha_m$ (\ref{eq2}) and $\iota_m$ (\ref{eq3}) in liquids is notable, as is the closeness of the lower bounds of $\alpha_m$ and $\nu_m$ despite the fundamental physical distinction between energy and momentum diffusion and very different ways of measuring $\alpha$ and $\nu$.

\section{Results and discussion}

\subsection{Derivation of the thermal diffusivity minimum}

In this section, we derive the thermal diffusivity at its minimum. We start our discussion with the thermal diffusivity due to ionic motion, and will comment on the electron conductivity later. We will see that Eqs. \eqref{eq2} and \eqref{eq3} emerge from connecting thermal diffusivity at the minimum to quantum-mechanical properties of condensed matter phases including the Bohr radius and Rydberg energy.

\begin{figure}[ht]
\begin{center}
{\scalebox{0.35}{\includegraphics{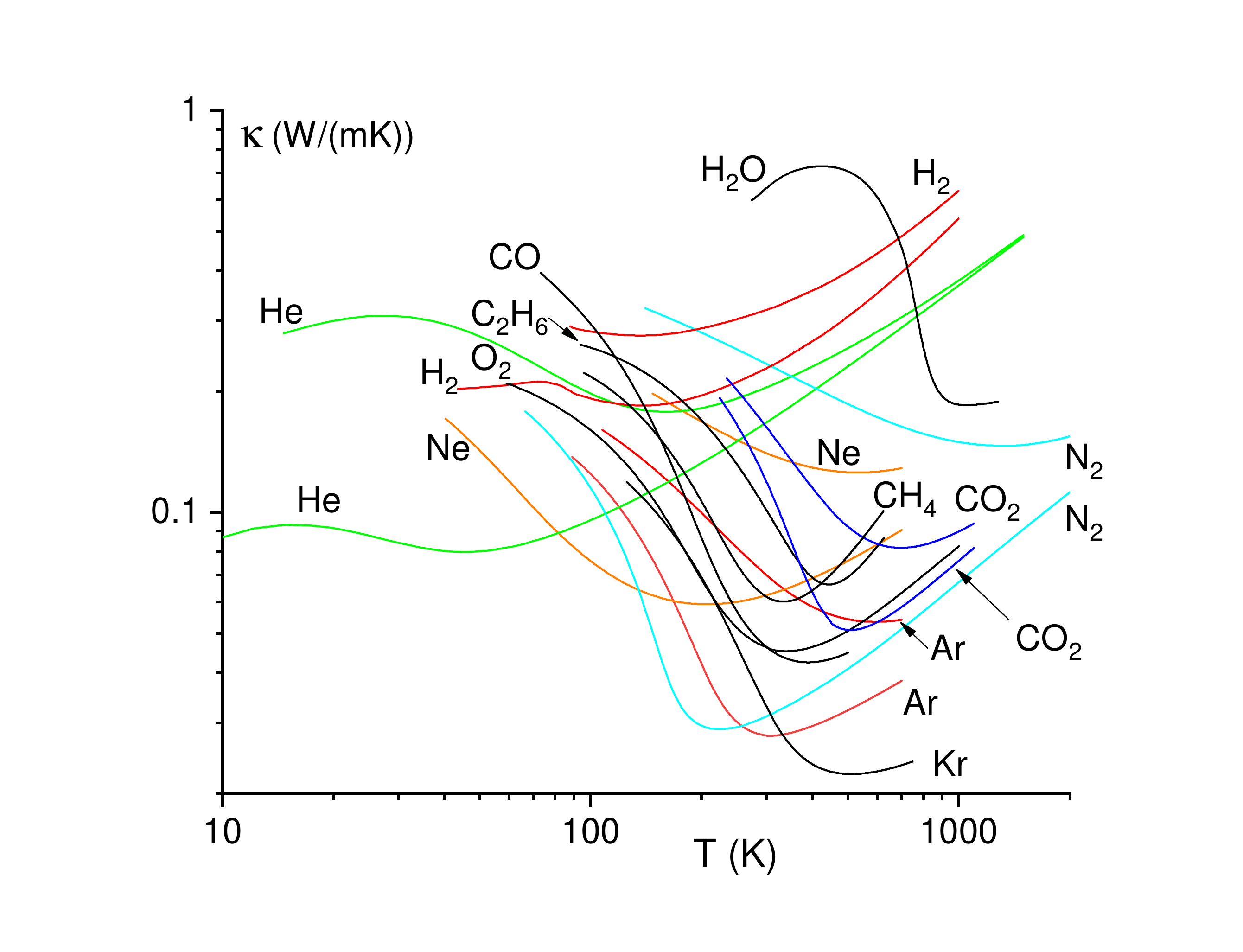}}}
{\scalebox{0.35}{\includegraphics{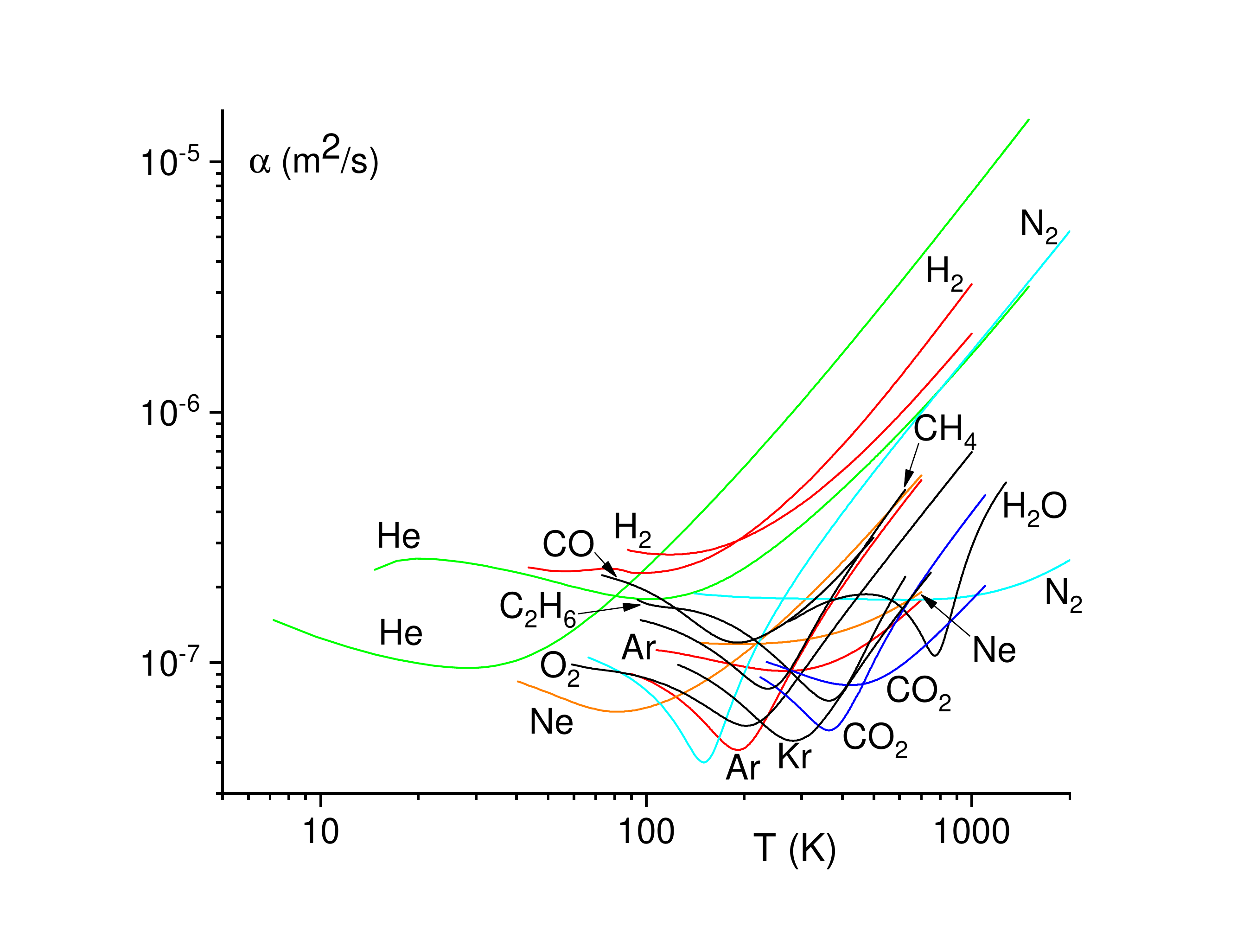}}}
\end{center}
\caption{Experimental thermal conductivity $\kappa$ (top) and thermal diffusivity $\alpha$ (bottom) of noble, molecular and network liquids \cite{nist} showing minima. $\kappa$ and $\alpha$ for Kr, O$_2$, H$_2$O, CH$_4$, C$_2$H$_6$ and CO are shown for pressure $P=30$ MPa, 30 MPa, 70 MPa, 20 MPa, 20 MPa and 20 MPa, respectively. $\kappa$ and $\alpha$ for Ar, Ne, He, N$_2$, $H_2$ and CO$_2$ are shown at two pressures each: 20 and 100 MPa for Ar, 50 and 300 MPa for Ne, 20 and 100 MPa for He, 10 MPa and 500 MPa for N$_2$, 50 MPa and 100 MPa for H$_2$, and 30 and 90 MPa for CO$_2$. The minimum at higher pressure is above the minimum at lower pressure for each fluid.}
\label{fig1}
\end{figure}

It is useful to first show the experimental data showing the minima. We have collected available experimental data \cite{nist} of $\kappa$ in several noble (Ar, Ne, He and Kr), molecular (N$_2$, H$_2$, O$_2$, CO$_2$, CH$_4$ C$_2$H$_6$ and CO) and network fluids (H$_2$O). Our selection includes industrially important supercritical fluids such as CO$_2$ and H$_2$O \cite{deben}. We have calculated $\alpha=\frac{\kappa}{\rho c_p}$ using the experimental values of $c_p$ and $\rho$ at respective temperatures and show both $\kappa$ and $\alpha$ in Fig. \ref{fig1}. For some fluids, we show the data at two different pressures. The low pressure was chosen to be far above the critical pressure so that the data are not affected by near-critical anomalies. The highest pressure was chosen to (a) make the pressure range considered as wide as possible and (b) be low enough in order to see the minima in the temperature range available experimentally. We observe that $\kappa$ and $\alpha$ universally have minima. We also observe that $\kappa$ can have weak maxima at low temperature related to the competition between the increase of heat capacity due to phonon excitations in the quantum regime and decrease of the phonon mean free path $l$ as in solids. In H$_2$O, the broad maximum is related to water-specific anomalies including broad structural transformation between differently-coordinated states.

We now move to the reason why $\kappa$ and $\alpha$ have minima in liquids as a function of temperature. In solids, the thermal conductivity $\kappa$ can be written as $\kappa=cvl$, where $c$ is the specific heat per volume unit \cite{ashcroft}, $v$ is the speed of sound, $l$ is the phonon mean free path and we dropped the numerical factor on the order of unity. Then, the diffusion constant is given by

\begin{equation}
\alpha=vl
\label{t1}
\end{equation}

In gases, $\alpha$ can be written in the same way as (\ref{t1}), but - and this reflects the difference between heat transfer in solids and gases - $v$ in (\ref{t1}) corresponds to the average velocity of gas molecules and $l$ to the molecule free path \cite{gases}.

The minimum of $\alpha$ is due to the crossover between the liquid-like and gas-like regimes of particles dynamics which we qualify below. Molecular motion in low-temperature liquids combines solid-like oscillations around quasi-equilibrium positions and diffusive jumps to new positions, enabling liquid flow. These jumps are due to temperature-induced molecular jumps over an energy barrier set by the interaction with other molecules, resulting in the exponential temperature dependence of viscosity. The jumps are characterised by liquid relaxation time, $\tau$, the average time between the molecular jumps \cite{frenkel}. The collective excitations in liquids (we refer to these as phonons or phonon-like modes in a wider sense \cite{ropp}) consist of one longitudinal mode and two transverse modes propagating above the threshold value in $k$-space \cite{ropp,yang}. The temperature increase has two effects on $\alpha$ in Eq.(\ref{t1}). First, the phonon mean free path $l$ decreases. Second, the speed of sound decreases as it does in solids. However, the decrease of $v$ and $l$ can not continue indefinitely due to the UV cutoff in condensed matter phases: $l$ is limited by the interatomic separation $a$ at the Mott-Ioffe-Regel (MIR) limit and $\tau$ is limited by the elementary vibration period, commonly approximated by the Debye vibration period $\tau_{\rm D}$.

An important effect related to reaching the UV cutoff is that further temperature increase results in the qualitative change of particle dynamics \cite{ropp,f1,f2}. On further temperature increase, the oscillatory component of molecular motion is lost, and molecules start moving in a purely diffusive manner. At high temperature and/or low density, molecules gain enough energy to move distance $l_p$ without collisions with velocity $v_t$, where $l_p$ is particle mean free path and $v_t$ is thermal velocity. $l_p$ and $v_t$ both increase with temperature. Therefore, $\alpha$ in Eq.(\ref{t1}) has a minimum. The same argument leading to a minimum applies to $\kappa=c\rho\alpha$. In the liquid-like regime, $\rho$ and $c$ are monotonically decreasing functions of temperature \cite{ropp}, hence the minima of $\alpha$ and $\kappa$ can take place at different temperature.

If the temperature is increased at pressure below the critical point, the system crosses the boiling line and undergoes the liquid-gas transition. As a result, $\alpha$ and $\kappa$ undergo a sharp change at the phase transition, rather than showing a smooth minimum as in Fig. \ref{fig1}. In order to avoid the effects related to the phase transition, we need to consider the supercritical state. Here, the Frenkel line \cite{ropp,f1,f2} formalises the qualitative change of molecular dynamics from combined oscillatory and diffusive to purely diffusive. The Frenkel line touches the boiling line slightly below the critical point and extends to arbitrarily high temperature and pressure on the phase diagram. At sufficiently high pressure and temperature, it runs nearly parallel to the boiling line in the logarithmic (pressure, temperature) coordinates \cite{f1}. The location of minima of different properties such as viscosity or thermal conductivity may depend on the path taken on the phase diagram. As a result, the minima may deviate from the Frenkel Line depending on the path \cite{ropp}.

Before evaluating $\alpha_m$, we first see how well we can estimate $\kappa$ at the minimum, $\kappa_m$, using our approach. When $l$ becomes comparable to $a$ at the minimum, the velocity $v$ can be evaluated as $v=\frac{a}{\tau_{\rm D}}$ because the time for a molecule to move distance $a$ in this regime is given by the characteristic time scale set by $\tau_{\rm D}$. Recalling that $c$ featuring in $\kappa=cvl$ is the temperature derivative of energy density \cite{ashcroft}, $c=\frac{c_v}{a^3}$, where $c_v$ is heat capacity per atom at constant volume (the derivative is taken at constant volume) and $a^{-3}$ is the concentration. At the minimum, $c_v$ is close to $2\,k_{\rm B}$, reflecting the disappearance of two transverse modes at the dynamical crossover \cite{ropp,yang}. Setting $l=a$, $v=\frac{a}{\tau_{\rm D}}=\frac{1}{2\pi}\omega_{\rm D}a$, where $\omega_{\rm D}$ is Debye frequency, gives

\begin{equation}
\kappa_m=\frac{1}{\pi}\frac{k_{\rm B}\,\omega_{\rm D}}{a}
\label{km}
\end{equation}

Taking the typical values of $a=$3-6 \AA\ and $\frac{\omega_{\rm D}}{2\pi}$ on the order of 1 THz, we find $\kappa_m$ in the range $0.05-0.09$ $\frac{{\rm W}}{{\rm mK}}$, providing an order of magnitude estimation of $\kappa_m$ consistent with the experimental minima in Fig. 1a. This sets the stage for our later calculation of thermal diffusivity at its minimum using fundamental physical constants.

We note that the minima of $\kappa$ in Fig. 1a is lower than thermal conductivity in low-$\kappa$ solids such as SnSe ($\kappa=0.23~\frac{{\rm W}}{{\rm mK}}$) where it is considered as ``ultralow'' \cite{nature}.

We also observe that high pressure reduces $a$ and increases $\omega_{\rm D}$. Eq. (\ref{km}) predicts that $\kappa_m$ increases with pressure as a result, in agreement with the experimental behavior in Fig.\ref{fig1}. We note that (\ref{t1}) applies in the regime where $l$ is larger than $a$, and in this sense our evaluation of conductivity minimum is an order-of-magnitude estimation, as are our other results below. In this regard, we note that theoretical models can only describe a dilute gas limit where perturbation theory applies \cite{gases}, but not in the regime where $l$ is comparable to $a$ and where the energy of inter-molecular interaction is comparable to the kinetic energy. In view of theoretical issues, we consider our evaluation useful. In addition to be informative, an order-of-magnitude evaluation is perhaps unavoidable if a complicated property such as thermal conductivity is to be expressed in terms of fundamental constants only.

We are now ready to evaluate $\alpha$ at its minimum, $\alpha_m$. As discussed above, $l$ at the minimum is $l\approx a$. The speed of sound $v$ in the Debye model is $v=\frac{a}{\tau_{\rm D}}$ (at the crossover where $\tau$ becomes comparable to the time it takes the molecule to move distance $a$ and where $\tau\approx\tau_{\rm D}$ as discussed above, $v$ becomes approximately equal to thermal velocity). Using $l=a$ and $v=\frac{a}{\tau_{\rm D}}=\frac{1}{2\pi}a\omega_{\rm D}$ in (\ref{t1}) gives

\begin{equation}
\alpha_m=\frac{1}{2\pi}\omega_{\rm D}a^2
\label{alpham}
\end{equation}

The energy diffusion constant $\alpha_m$ in (\ref{alpham}) can now be related to fundamental physical constants by recalling that the properties defining the UV cutoff in condensed matter can be expressed in terms of fundamental constants \cite{sciadv}. For the benefit of the reader and later discussion, we reproduce the brief derivation below. Two relevant quantities are Bohr radius, $a_{\rm B}$, setting the characteristic scale of inter-particle separation on the order of Angstrom:

\begin{equation}
a_{\rm B}=\frac{4\pi\epsilon_0\hbar^2}{m_ee^2}
\label{bohr}
\end{equation}

\noindent and the Rydberg energy, $E_{\rm R}=\frac{e^2}{8\pi\epsilon_0a_{\rm B}}$ \cite{ashcroft}, setting the characteristic scale for the cohesive energy in condensed matter phases on the order of several eV:

\begin{equation}
E_{\rm R}=\frac{m_ee^4}{32\pi^2\epsilon_0^2\hbar^2}
\label{rydberg}
\end{equation}

\noindent where $e$ and $m_e$ are electron charge and mass.

The characteristic phonon energy $\hbar\omega_{\rm D}$ is related to the cohesive energy $E$, $\frac{\hbar\omega_{\rm D}}{E}$ as:

\begin{equation}
\frac{\hbar\omega_{\rm D}}{E}=\left(\frac{m_e}{m}\right)^{\frac{1}{2}}
\label{ratio}
\end{equation}

\noindent which, up to a factor close to 1, follows from approximating $\hbar\omega_{\rm D}$ as $\hbar\left(\frac{E}{ma^2}\right)^{\frac{1}{2}}$, taking the ratio $\frac{\hbar\omega_{\rm D}}{E}$ and using $a=a_{\rm B}$ from (\ref{bohr}) and $E=E_{\rm R}$ from (\ref{rydberg}).

Combining (\ref{alpham}) and (\ref{ratio}) gives

\begin{equation}
\alpha_m=\frac{1}{2\pi}\frac{E a^2}{\hbar}\left(\frac{m_e}{m}\right)^{\frac{1}{2}}
\label{nu01}
\end{equation}

The parameters $a$ and $E$ in (\ref{nu01}) are set by their characteristic scales $a_{\rm B}$ and $E_{\rm R}$ as discussed earlier. Using $a=a_{\rm B}$ from (\ref{bohr}) and $E=E_{\rm R}$ from (\ref{rydberg}) in (\ref{nu01}) gives a remarkably simple equation for $\alpha_m$ as in Eq. \eqref{eq2}, which we reproduce below for convenience:

\begin{equation}
\alpha_m=\frac{1}{4\pi}\frac{\hbar}{\sqrt{m_em}}
\label{nu1}
\end{equation}

Eq. (\ref{nu1}) is the main result of this paper. The same result for $\alpha_m$ in (\ref{nu1}) can be obtained without explicitly using $a_{\rm B}$ and $E_{\rm R}$ in (\ref{nu01}). The cohesive energy, or the characteristic energy of electromagnetic interaction, is

\begin{equation}
E=\frac{\hbar^2}{2m_ea^2}
\label{energy}
\end{equation}

Using (\ref{energy}) in (\ref{nu01}) gives (\ref{nu1}).

We now analyze (\ref{nu1}) and its implications. $\alpha_m$ contains $\hbar$ and electron and molecule masses only. $m$ characterises the molecules involved in heat transfer. $m_e$ characterises electrons setting the inter-molecular interactions. The quantum origin of $\alpha_m$, signified by $\hbar$ in (\ref{nu1}), is due to the quantum nature of inter-particle interactions.

The mass $m$ in (\ref{nu1}) is $m=Am_p$, where $A$ is the atomic weight and $m_p$ is the proton mass. The inverse square root dependence $\alpha_m\propto\frac{1}{\sqrt{A}}$ interestingly implies that for different liquids $\alpha_m$ varies by a factor of about 10 only. Setting $m=m_p$ ($A=1$) for H in (\ref{nu1}) (similarly to (\ref{bohr}) and (\ref{rydberg}) derived for the H atom) gives the fundamental thermal diffusivity in terms of $\hbar$, $m_e$ and $m_p$ as
\begin{equation}
\alpha_m=\frac{1}{4\pi}\frac{\hbar}{\sqrt{m_em_p}}\approx 10^{-7} \frac{\rm{m}^2}{\rm{s}}
\label{nuf}
\end{equation}

For the lightest element, H, Eq. (\ref{nuf}) gives the maximal value of $\alpha_m$. It is interesting to ask what quantity has an absolute minimum. If we define the ``elementary conductivity'' $\iota$ (``iota'') equivalent to the elementary viscosity \cite{sciadv} as $\iota=\alpha_m m$, Eq. (\ref{nu1}), gives $\iota=\frac{\hbar}{4\pi}\left({\frac{m}{m_e}}\right)^{\frac{1}{2}}$. $\iota$ has the absolute minimum, $\iota_m$, for H where $m$ is the proton mass $m_p$:
\begin{equation}
\iota_m=\frac{\hbar}{4\pi}\left({\frac{m_p}{m_e}}\right)^{\frac{1}{2}}
\label{iota}
\end{equation}

\noindent and is on the order of $\hbar$.

Eq. (\ref{iota}) interestingly involves the proton-to-electron mass ratio, one of few dimensionless combinations of fundamental constants of importance in a variety of areas \cite{barrow}. Together with the fine structure constant, this ratio has a particular importance from the point of view of governing nuclear reactions, synthesis in stars and creation of planets and heavier elements including carbon. The balance between the two dimensionless constants provides a narrow ``habitable zone'' where stars and planets can form and life-supporting molecular structures can emerge \cite{barrow}.

\subsection{Comparison to the experimental data}

We now compare our bounds to experiments. In Table \ref{tab1} we compare $\alpha_m$ calculated according to (\ref{nu1}) to the experimental $\alpha_m$ \cite{nist} for all liquids shown in Fig. \ref{fig1}. The ratio between experimental and predicted $\alpha_m$ is in the range of about $0.9-4$. The ratio is the largest for fluids under high pressure (e.g. N$_2$ at 500 MPa and Ar at 100 MPa) which our Eq. (\ref{nu1}) does not account for as discussed below. For the lightest liquid, H$_2$, experimental $\alpha_m$ is close to the theoretical fundamental thermal diffusivity viscosity (\ref{nuf}). We therefore find that (\ref{nu1}) predicts the right order of magnitude of $\alpha_m$.

\begin{table}[ht]
\begin{tabular}{ l| c c c c}
                   & $\alpha^{th}_m=\nu^{th}_m$     & $\alpha^{exp}_m$       & $\nu^{exp}_m$ &$\nu_m/\alpha_m$\\
\hline\\
Ar (20 MPa)        & 3.4                   & 4.5                & 5.9  & 1.3\\
Ar (100 MPa)       & 3.4                   & 9.3                & 7.7  & 0.8\\
Ne (50 MPa)        & 4.8                   & 6.4                & 4.6  & 0.7\\
Ne (300 MPa)       & 4.8                   & 11.9               & 6.5  & 0.6\\
He (20 MPa)        & 10.7                  & 9.5                & 5.2  & 0.6\\
He (100 MPa)       & 10.7                  & 17.9               & 7.5  & 0.4\\
Kr (30 MPa)        & 2.3                   & 4.9                & 5.2 &1.1\\
N$_2$ (10 MPa)     & 4.1                   & 4.0                & 6.5  &1.6\\
N$_2$ (500 MPa)    & 4.1                   & 17.8               & 12.7 & 0.7\\
H$_2$ (50 MPa)     & 15.2                  & 22.8               & 16.3 & 0.7\\
H$_2$ (100 MPa)    & 15.2                  & 27.0               & 19.4 &0.7\\
O$_2$ (30 MPa)     & 3.8                   & 5.6                & 7.4  &1.3\\
H$_2$O (70 MPa)    & 5.1                   & 10.7               & 11.9 &1.1\\
CO$_2$ (30 MPa)    & 3.2                   & 5.4                & 8.0  &1.5\\
CO$_2$ (90 MPa)    & 3.2                   & 8.1                & 9.3  &1.2\\
CH$_4$ (20 MPa)    & 5.4                   & 7.9                & 11.0 &1.4\\
C$_2$H$_6$ (20 MPa)& 3.9                   & 7.0                & 12.0 &1.7\\
CO (20 MPa)        & 4.1                   & 12.0               & 7.7  &0.6\\
\label{table}
\end{tabular}
\caption{Theoretical (th) and experimental (exp) values for the thermal diffusivity $\alpha_m$ and the kinematic viscosity $\nu_m$ at the minima. All the quantities are displayed in units of $\times$10$^8$ m$^2$/s except from the last ratio which is dimensionless.}
\label{tab1}
\end{table}

We observe that $\alpha$ increases with pressure in Table \ref{tab1}, similarly to $\kappa$ in Fig.\ref{fig1}. However, the pressure dependence is not accounted in $\alpha_m$ in (\ref{nu1}) since (\ref{nu1}) is derived in the approximation involving Eqs. (\ref{bohr})-(\ref{nu01}) which do not account for the pressure dependence of $\omega_{\rm D}$ and $E$.

We make three further remarks regarding the comparison of theoretical and experimental results in Table \ref{tab1}. First, the important term in Eq. (\ref{nu1}) includes the combination of fundamental constants which sets the characteristic scale of the lower bound of thermal diffusivity, whereas the numerical factor in (\ref{nu1}) may be affected by the approximations used as discussed earlier. Second, Eqs. (\ref{bohr})-(\ref{ratio}) assume valence electrons setting strong bonding such as covalent and ionic. Thermal conductivity of these systems in the supercritical state is unavailable due to high critical points. The available data \cite{nist} used in Fig. \ref{fig1} and Table \ref{tab1} include weakly-bonded systems such as molecular, noble and hydrogen-bonded fluids. Bonding in these systems is also electromagnetic in origin, although weak van der Waals and dipole interactions result in smaller $E$ and, therefore, smaller $\alpha$. However, we note that the dependence of $\alpha_m$ on bonding type is weak because (a) $\alpha_m$ in (\ref{nu01}) contains the factor $Ea^2$ and (b) $a$ is 2-4 times larger and $E^{\frac{1}{2}}$ is 3-10 times smaller in weakly-bonded  as compared to strongly-bonded systems \cite{vadim1}. As a result, the order-of-magnitude evaluation (\ref{nu1}) is unaffected as Table \ref{fig1} shows. Third, Eq. (\ref{nu1}) for strongly-bonded nonmetallic (covalent and ionic) fluids gives a {\it prediction} for future experimental work.

The lower bound setting $\alpha_m$ in (\ref{nu1}) is consistent with the uncertainty principle. As discussed earlier, the minimum of $\alpha$ can be evaluated as $\alpha_m=va=\frac{pa}{m}$, where $p$ is particle momentum. Using the uncertainty relation applied to a particle localised in the region set by $a$, $\alpha_m\ge\frac{\hbar}{m}$. $\frac{\hbar}{m}$ is smaller than $\alpha_m$ in (\ref{nu1}) by the factor $F=\frac{1}{4\pi}\left({\frac{m}{m_e}}\right)^{\frac{1}{2}}$. $F\approx 22$ in Ar and becomes smaller for lighter systems. Therefore, the minimum (\ref{nuf}) provides a stronger bound as compared to the uncertainty relation. 

An important difference of our lower bound (\ref{nu1}) and bounds based on the uncertainty relations in earlier discussions \cite{zaanen2,hartnoll1,behnia1,behnia2,kss,hartnoll2,spin1,zaanen1} is that (\ref{nu1}) corresponds to a true minimum of thermal diffusivity as seen in Fig. 1 (in a sense that the function has an extremum), whereas the uncertainty relation compares a product ($px$ or $Et$) to $\hbar$ but the product does not necessarily correspond to a minimum of a function and can apply to a monotonic function.

\subsection{Energy and momentum diffusion}

We now discuss the relationship between the minima of $\alpha$ and the minima of kinematic viscosity $\nu$, $\nu_m$.

Interestingly, the question of viscosity minima was raised before. Purcell observed \cite{purcell} that ``viscosities have a big range but they stop at the same place.'' In the earlier work, we have ascertained the lower limit of kinematic viscosity in terms of fundamental constants \cite{sciadv}.

We plot the experimental $\alpha$ and $\nu$ for two noble and two molecular liquids in Fig. 2 at the same pressure as in Fig. 1. We observe the closeness of the minima of both properties. This is unexpected and is surprising, in view that the two properties are physically distinct and are measured very differently. We compare $\alpha_m$ and $\nu_m$ below in detail.

\begin{figure}[ht]
\begin{center}
{\scalebox{0.35}{\includegraphics{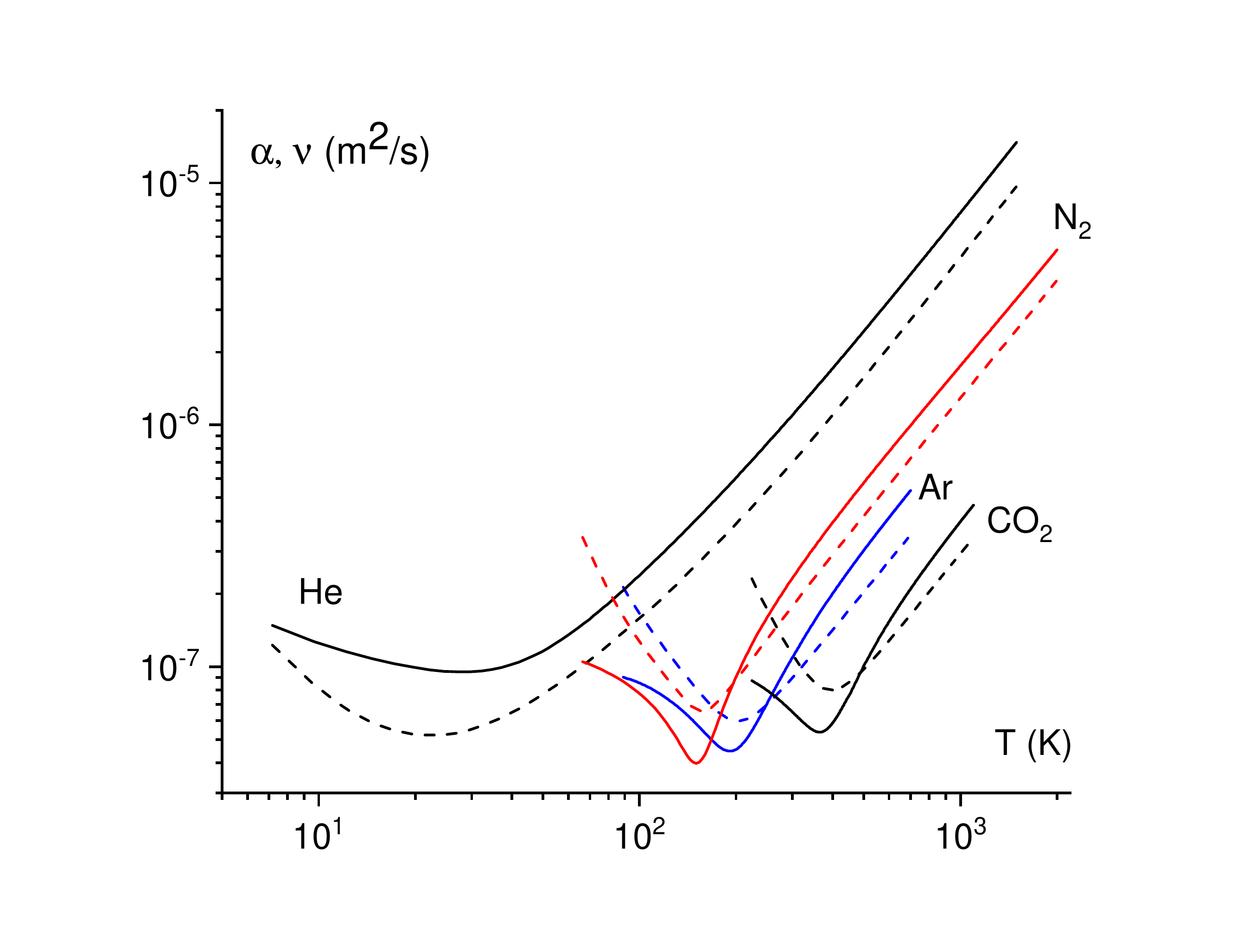}}}
\end{center}
\caption{Experimental $\alpha$ (solid lines) and $\nu$ (dashed lines) for He (20 MPa), N$_2$ (10 MPa), Ar (20 MPa) and CO$_2$ (30 MPa) \cite{nist}.}
\label{fig2}
\end{figure}

There are interesting and important similarities and differences between the two properties. The first analogy is that Eq. (\ref{Heat}), which describes energy diffusion, is analogous to that determining momentum diffusion if $T$ is replaced by the velocity field and $\alpha$ is replaced by $\nu$. Second, recall that the minimum of thermal conductivity is due to $v$ and $l$ changing from the phonon speed and phonon mean free path in the low-temperature liquid-like regime to particle thermal speed and particle mean path in the high-temperature gas-like regime. The minimum of liquid viscosity is due to the crossover between the exponential decrease of viscosity in the low-temperature liquid-like regime $\eta\propto\exp\left(\frac{U}{T}\right)$ to $\eta\propto\rho v l$ in the high-temperature gas-like regime, where $U$ is the activation barrier for diffusive particle rearrangements, $v$ and $l$ are particle thermal speed and mean free path, respectively. Therefore, the temperature dependence of the thermal conductivity and the viscosity is the same in the gas-like regime at high temperature but is different in the liquid-like regime at low temperature. Third and finally, the dominant contribution to thermal conductivity in the low-temperature liquid-like regime is due to phonons as in solids. In the high-temperature gas-like regime, thermal conductivity is due to particle collisions. Viscosity, on the other hand, is due to the dynamics of individual particles and momentum they transfer in both liquid-like and gas-like regimes. Therefore, thermal conductivity and viscosity are set by the same process at high temperature but by different processes at low. Consistent with this picture, Fig. \ref{fig2} shows that temperature behavior of $\alpha$ and $\nu$ is more similar at high temperature as compared to low.

Despite the above differences between $\alpha$ and $\nu$, theoretical values at their {\it minima} are the same. Indeed, we have previously shown \cite{sciadv} that the minima of $\nu$, $\nu_m$, are given by Eq. (\ref{alpham}), or Eq. (\ref{nu1}) involving fundamental physical constants, implying

\begin{equation}
\nu_m=\alpha_m
\label{equal}
\end{equation}

Therefore, the closeness between $\nu_m$ and $\alpha_m$ is explained by observing that both $\alpha$ and $\nu$ at their minima are governed by UV properties such as Bohr radius and Debye frequency in Eq. (\ref{alpham}).

We have calculated $\nu=\frac{\eta}{\rho}$ using the experimental values of viscosity $\eta$ and density $\rho$ \cite{nist} for all liquids at the same pressure as thermal conductivity in Fig. \ref{fig1} and show the minima of $\nu$, $\nu_m$ in the third column in Table \ref{tab1}. We observe that the experimental values of $\alpha_m$ and $\nu_m$ are close to each other. This agreement is also seen in the last column of Table \ref{tab1} where the ratio $\nu_m/\alpha_m$ is in the range 0.4-1.7. We note that the temperatures of the minima of $\alpha_m$ and $\nu_m$ are somewhat different, nevertheless the closeness of $\alpha_m$ and $\nu_m$ implies that the Prandtl number, $\frac{\nu}{\alpha}$, is on the order of 1 at temperatures close to the minima. This is seen in the last column of Table \ref{tab1}.

The agreement between experimental $\alpha_m$ and $\nu_m$ as well as their agreement with the theoretical estimation in the first column in Table \ref{tab1} importantly reinforces our analysis of the minima and adds to its consistency.

\begin{figure}[ht]
\begin{center}
{\scalebox{0.35}{\includegraphics{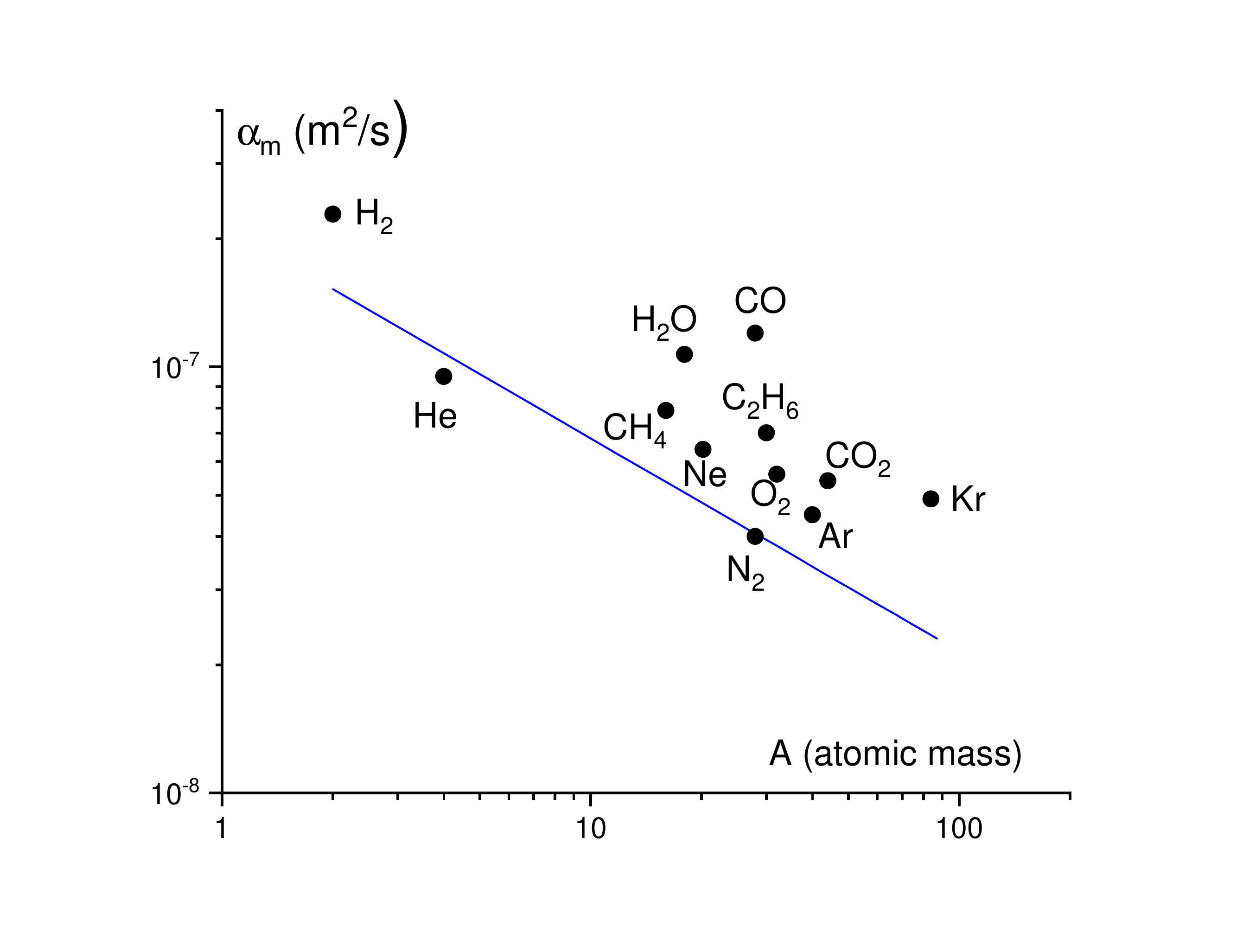}}}
{\scalebox{0.35}{\includegraphics{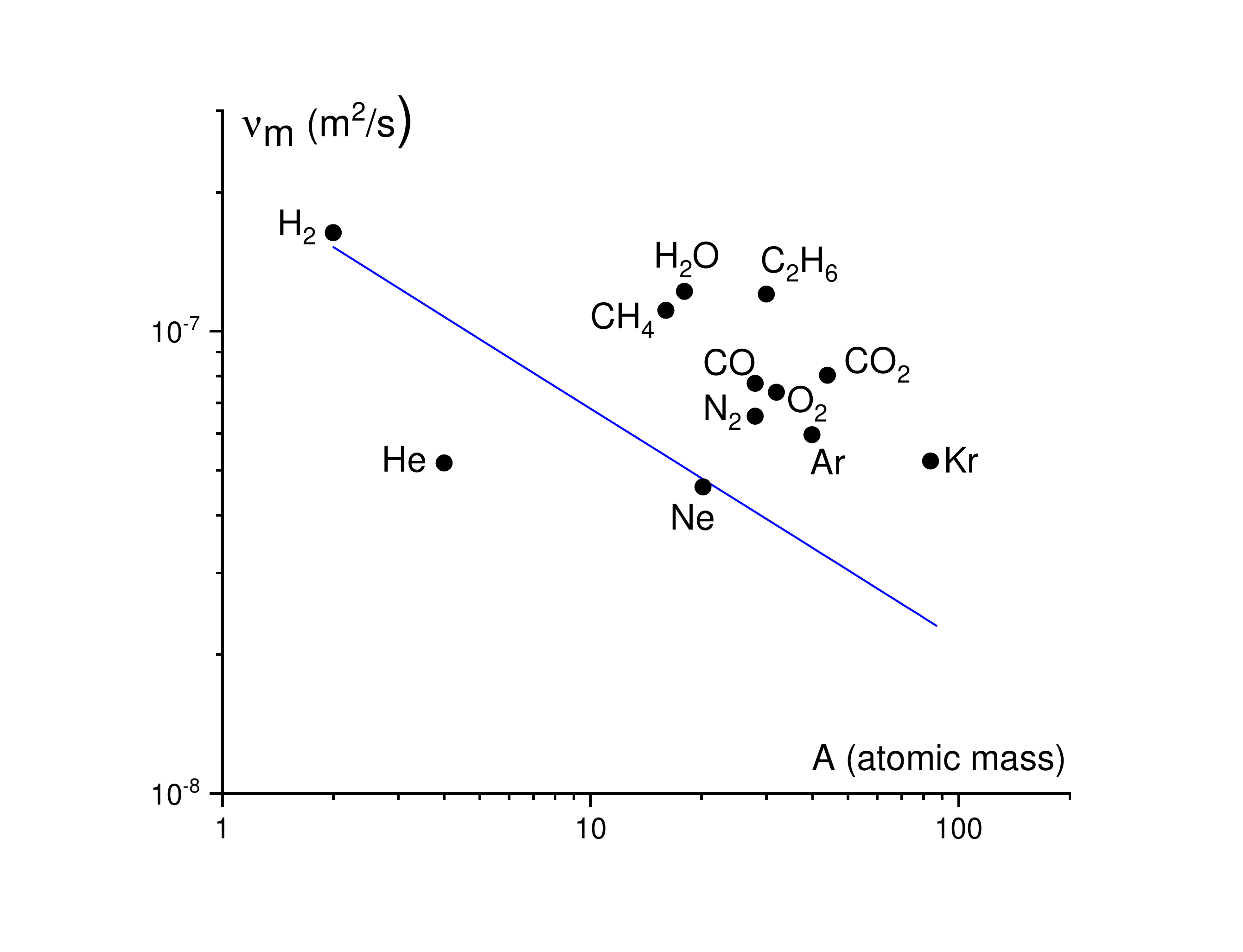}}}
\end{center}
\caption{Points show experimental $\alpha_m$ (top) and $\nu_m$ (bottom) as a function of molecular mass. The solid line is the prediction of Eq. \eqref{nu1}. 
}
\label{mass}
\end{figure}

Our final comparison of the theoretical result and experimental data concerns the inverse square-root dependence of $\alpha_m$ and $\nu_m$: according to Eqs. \eqref{nu1} and \eqref{equal}, $\alpha_m,\nu_m\propto\frac{1}{\sqrt{m}}$. Fig. \ref{mass} shows the experimental $\alpha_m$ and $\nu_m$ of all systems in Table 1 at low pressure as a function of the molecule mass, together with the solid line representing the theoretical result \eqref{nu1}. We observe a trend of both $\alpha_m$ and $\nu_m$ reducing with molecular mass. We also observe that nearly all experimental plots are above the theoretical prediction of the lower bound. 
We note that the inverse square-root dependence is expected for strong electromagnetic interactions where energy and interatomic distance do not depend on the ion mass. For weak interactions, the energy depends on the size of the atom or molecule \cite{vadim1}. This contributes to the scatter of points in Fig. \ref{mass}.

We note that the above discussion applies to systems where the dominant contribution to thermal diffusivity is related to the motion of ions rather than electrons (the electron mass $m_e$ enters Eq. \eqref{nu1} because $m_e$ enters the Bohr radius \eqref{bohr} and Rydberg energy \eqref{rydberg}. The minima of $\alpha$ due to electrons will be discussed elsewhere. Here, we note that thermal conductivity of both high-temperature solid and liquid metals is typically in the range 10-100 $\frac{\rm W}{\rm mK}$ and 2-3 orders of magnitude higher than in insulators \cite{drits} due to the electronic contribution (this is related to smaller electron mass compared to ion mass.) Hence the minimum discussed here applies to conducting systems too.

\subsection{Minima on the phase diagram}

$\alpha_m$ provides a useful guidance for the minimal value of thermal diffusivity achieved for a given material. This can be important, for example, in the area of thermal insulation. Small values of thermal conductivity are also important in other areas such as enhancing the thermoelectric effect. As already noted, the exceptionally low thermal conductivity reported in Ref. \cite{nature} for the solid with high thermoelectric figure is still larger than the minima of $\kappa$ in Fig. 1a.

It is interesting to ask whether the minima of $\nu_m$ and $\alpha_m$ discussed for the liquid and supercritical states apply to other parts of the phase diagram. In solids, $\alpha=vl$ in Eq. \eqref{t1} is larger because (a) the speed of sound $v$ is faster and (b) the mean free path $l$ is larger than that in liquids and is typically larger than $a$ at the UV cutoff. It can be seen that $vl$ similarly increases in gases if we recall that the minima at the UV cutoff approximately correspond to the Frenkel line \cite{ropp,f1,f2}. The speed of sound is approximately equal to the thermal speed of particles at the line and increases above the line in the gas-like state as thermal velocity $\propto\sqrt{T}$. $l$ becomes the particle mean free path above the line in the gas-like state and similarly increases with temperature. Hence, $\alpha=vl$ increases in gases, and the minimum of $\alpha$, $\alpha_m$ at the UV cutoff, applies to all three states of matter.

The minima $\alpha_m$ and $\nu_m$ behave differently in close proximity to the critical point. Indeed, viscosity diverges at the critical point \cite{xenon-visc}, and $\nu_m$ increases close to the critical point. Therefore, $\nu_m$ gives the global minimum on the entire phase diagram. On the other hand, isobaric heat capacity diverges much faster than $\kappa$ \cite{anisimov}, and $\alpha$ at the critical point tends to zero as a result. Therefore, $\alpha_m$ gives the minimum on the phase diagram except in the vicinity of the critical point.

\section{Conclusions}

In summary, we have shown that thermal diffusivity of liquids and supercritical fluids has a lower bound which is fixed by fundamental physical constants for each fluid. The newly introduced elementary thermal diffusivity has an absolute lower bound dependent on $\hbar$ and the proton-to-electron mass ratio only. We have also shown that (a) the lower bound of thermal diffusivity theoretically coincides with the lower bound of kinematic viscosity and (b) the ratio between experimental minima of the two properties is close to 1. This finding implies a universal lower bound for two distinct properties, energy and momentum diffusion which, to the best of our knowledge, has not been discussed before.\\

We are grateful to J. Zaanen and S. Hartnoll for discussions. M.B. acknowledges the support of the Spanish MINECO’s ``Centro de Excelencia Severo Ochoa'' Programme under grant SEV-2012-0249. K. T. acknowledges the EPSRC support.




\end{document}